\documentstyle[11pt,psfig,amssymb]{article}
\begin{document}
\newcommand{\be}{\begin{equation}}
\newcommand{\ee}{\end{equation}}
\newcommand{\lb}{\label}
\newcommand{\en}{\epsilon}
\newcommand{\ven}{\varepsilon}
\newcommand{\bu}{{\bf u}}
\newcommand{\bv}{{\bf v}}
\newcommand{\bx}{{\bf x}}
\newcommand{\bk}{{\bf k}}
\newcommand{\bn}{{\bf n}}
\newcommand{\bE}{{\bf e}}
\newcommand{\bF}{{\bf f}}
\newcommand{\bpsi}{{\mbox{\boldmath $\psi$}}}

\textwidth6.25in
\textheight8.5in
\oddsidemargin.25in
\topmargin0in

\title{Gibbsian Hypothesis in Turbulence}
\author{Gregory L. Eyink$^{1,2}$, Shiyi Chen$^{2,3,4,5}$ and Qiaoning
Chen$^{3}$\\
$\,\,$\\
${}^{1}$Department of Mathematics, \\
University of Arizona, \\
Tucson, AZ 85721\\
$\,\,$\\
${}^{2}$Department of Mathematical Sciences \\
and \\
${}^{3}$Department of Mechanical Engineering, \\
The Johns Hopkins University, \\
Baltimore, MD 21218\\
$\,\,$\\
${}^{4}$Center for Nonlinear Studies and Theoretical Division, \\
Los Alamos National Laboratory, \\
Los Alamos, NM 87545\\
$\,\,$\\
${}^{5}$Peking University, China\\
}
\date{}
\maketitle
\begin{abstract}
We show that Kolmogorov multipliers in turbulence cannot be statistically
independent of others at adjacent scales (or even a finite range apart)
by numerical simulation of a shell model and by theory. As the simplest
generalization of independent distributions, we suppose that the
steady-state statistics of multipliers in the shell model are given
by a translation-invariant Gibbs measure with a short-range potential,
when expressed in terms of suitable ``spin'' variables: real-valued spins
that are logarithms of multipliers and XY-spins defined by local dynamical
phases. Numerical evidence is presented in favor of the hypothesis
for the shell model, in particular novel scaling laws and derivative
relations predicted by the existence of a thermodynamic limit.
The Gibbs measure appears to be in a high-temperature, unique-phase
regime with ``paramagnetic'' spin order.

\end{abstract}

\newpage


\section{Introduction}

In a famous paper on the local structure of turbulence of incompressible fluids
\cite{K62},
A. N. Kolmogorov in 1962 considered inertial-range multipliers defined  by
ratios of velocity
increments $w_{ij}(\ell,\ell')=\delta_{i,\ell}v_j/\delta_{i,\ell'}v_j.$ Here
$\delta_{i,\ell}v_j(\bx)
=v_j(\bx+\ell \bE_i)-v_j(\bx)$ is the increment of the $j$th component of the
velocity vector $\bv$
along the unit vector $\bE_i$. Kolmogorov hypothesized that, at very high
Reynolds number, these
multipliers should have distributions which are universal functions only of the
scale-ratio $\ell/\ell'$
and not of the absolute scale. He postulated further that multipliers
corresponding to widely
separated scales should be statistically independent. See also \cite{Kr74}. R.
Benzi, L. Biferale,
and G. Parisi \cite{BBP} have made a more precise and quite remarkable
hypothesis in the context of a
``shell model'' of turbulence. The latter are quadratically nonlinear dynamical
systems for
variables in a finite number $N$ of shells with wavenumbers $k_n=\lambda^n
k_0,\,\,n=1,...,N$ with
$\lambda > 1$.
For example, the SABRA model \cite{LPPPV} has one complex mode $u_n$ per shell
which obeys a
dynamical equation
\begin{eqnarray}
 {{du_n}\over{dt}} & = & i(ak_{n+1}u_{n+2}u_{n+1}^*+bk_nu_{n+1}u_{n-1}^* \cr
 \, & &\,\,\, -ck_{n-1}u_{n-1}u_{n-2})-\nu k_n^2 u_n + f_n. \lb{SABRA}
\end{eqnarray}
(\cite{BBP} considered a slightly different model.) Here $*$ stands for complex
conjugation,
$f_n$ is a forcing term which is restricted to the first few shells and $\nu$
is the ``viscosity''.
When $a+b+c=0$, equation (\ref{SABRA}) satisfies conservation of ``energy''
$E={{1}\over{2}}\sum_n |u_n|^2$
and ``helicity'' $H=\sum_n (a/c)^n |u_n|^2$ in the $\nu\rightarrow 0$ limit,
analogous to the quadratic
invariants of the inviscid Euler equations. Then, for large $N$ and high
Reynolds number
$Re=\sqrt{\langle|u_1|^2\rangle}/(\nu k_0)$, the authors of \cite{BBP}
hypothesized that the variables
$u_n,\,\,n=1,...,N$ should have a steady-state statistical distribution given
by a {\it Gibbs measure}
\be P(u_1,...,u_N) \propto \exp[-\sum_n \Phi_n(u_n,u_{n-1},u_{n+1},...)].
                                                  \lb{Gibbs} \ee
Because turbulence is a dissipative state, far from thermodynamic equilibrium,
the potentials
$\Phi_n$ have nothing to do with the inviscid invariants, even as
$\nu\rightarrow 0.$
For shellnumbers $n$ in the long inertial range, $1\ll n\ll N$, \cite{BBP}
supposed that the potential
$\Phi_n$ becomes a universal function $\Phi$ independent of $n$. They supposed
further that the potential
$\Phi$ is a sufficiently short-range function of ``the ratios between the $u$'s
and their angles''
\cite{BBP}. The authors of \cite{BBP} used an ``infinite-temperature'' model
with independent multipliers
to predict the scaling exponents.
It is usually assumed that such an approximation is qualitatively correct.
However, we show that the
multipliers cannot be strictly independent and that the correlations are
essential. We also give
evidence for the Gibbs hypothesis, based on theoretical analysis and direct
numerical simulation of
the shell-model dynamics (\ref{SABRA}).

\section{Theoretical Considerations}

First, let us give a more precise form to the hypothesis. In the shell model,
we introduce an
amplitude $\rho_n$ and a phase $\theta_n$ for each shellnumber $n$, via the
polar decomposition
$u_n= k_n^{-1/3}\cdot \rho_n e^{i\theta_n}$. Following \cite{BBP}, we separate
out the Kolmogorov
1941 scaling factor $k_n^{-1/3}$. If $w_n=\rho_n/\rho_{n-1}$ is the multiplier
defined by \cite{BBP}
and $\Delta_n = -\theta_n + \theta_{n-1}+\theta_{n-2}$ is the dynamical phase
factor \cite{BBP,LPPPV}, then we define
\be   \sigma_n=\ln(w_n),\,\,U_n=\exp(i\Delta_n) \lb{spins} \ee
where $\sigma_n$ is a local ``slope'' and $U_n$ is an ``XY-spin'' or
2-dimensional rotator spin. The definition of $\sigma_n$ is motivated by the
observation that $\ln \rho_n
= \sum_{k=1}^{n} \sigma_{k}$ is then a ``total spin''. Alternatively,
$h_n=\ln\rho_n$ can be viewed
as a ``height function'', as in equilibrium models of surface roughness. Then,
our hypothesis
is that the distribution of these ``spin variables''
$\xi_n=(\sigma_n,U_n)$ is a translation-invariant Gibbs measure with
a short-range potential $\Phi$, ignoring
finite-size effects from the forcing and dissipation ranges
of shellnumber $n$. The potential is expected to be, at least,
absolutely summable (\cite{Georgii}, section 2.1), which guarantees
its uniqueness up to physical equivalence. However, the numerical evidence
presented below suggests that the interactions are not merely summable,
but indeed quite rapidly decaying.

On the other hand, the potentials cannot have a strictly finite range $R$,
i.e. vanishing for any set of spin variables containing pairs $\xi_n,\xi_{n'}$
with $|n-n'|>R$. In particular, the assumption of zero-range interactions,
$R=0$, or independent spins which was made made by \cite{BBP} is ruled out.
Using exact constraints from the dynamics, we show in Appendix 1 that the
assumption of independent spins leads to the deterministic K41 fixed-point
$u_n \sim -i (\varepsilon/k_n)^{1/3}$ as the only statistically stable
solution.
In reality this solution is dynamically unstable \cite{SKL}. The false
assumption
of independence stabilizes this solution and prevents intermittency corrections
from developing. This is plausible, since intermittency is known to arise
in the shell models from ``burst'' solutions which exhibit long-range coherence
in both $\sigma_n$ and $U_n$ over many shells (e.g. see Fig.4 in
\cite{Gil-Domb}).
Furthermore, the K41 fixed-point can be shown to be stabilized by assuming {\it
any}
finite-range potential between spins \cite{ECC}, so that the stationary measure
cannot be Gibbs with any potential of strictly finite-range.

Another argument against independence of spins is that this would imply
the quadratic equation
\be a \mu_3^2 + b\mu_3 + c=0   \lb{4-5law} \ee
for the multiplier moment $\mu_p=\langle w^p\rangle,\,\,p=3$. Cf. \cite{BBP},
Eq.(18). This is analogous to the ``4/5-law'' of fluid turbulence . Note that,
in the independent spin approximation, the structure-function scaling exponents
are given by $\zeta_p =(p/3)-\log_\lambda\mu_p$. There are two roots of the
quadratic equation, $\mu_3=1$ and $\mu_3=c/a$. The first solution gives
constant
mean energy flux but zero helicity flux, while the second gives the
opposite. Thus, no joint cascade of energy and helicity is possible in the
independent spin approximation, contrary to observations \cite{Dit-Gul}.
Furthermore, the second solution violates the realizability inequality
$\mu_3\geq 0,$ when $c/a<0$ and the second invariant is truly
``helicity-like''.
These conclusions do not remain true for a potential of finite but non-zero
range.
The quadratic equation still holds for a nearest-neighbor potential or Markov
chain approximation, where now $\mu_3$ is the principal eigenvalue of a
``transfer matrix'' \cite{ECC}. As shown in Appendix 2 of the present paper,
joint cascades of energy and helicity are permitted in a Markov chain model,
if $\mu_3=1$ and the subleading eigenvalue $\mu_3'= c/a$. Furthermore, a
concrete
Markov chain model is constructed in the appendix to show that this situation
may be realized. While the Gibbs measure cannot be exactly nearest-neighbor,
we believe that this may be a good working approximation.

\section{Numerical Results}

We now present our simulation results for the SABRA model, with standard choice
of parameters $\lambda=2,
k_0 = 2^{-4}, a=1,b=c=-1/2$. We performed two sets of simulations, one with
$N=22,\,\nu=10^{-7}$ and
the second with $N=26,\,\nu=2\cdot 10^{-9}$. A force $f_n=F_n(1+i)/u_n^*$ was
used with $F_n$ real,
nonzero only for $n=2,3$, and chosen to give an input of energy but not
helicity. Except where stated,
the results shown are for the $N=26$ simulation. Stationary time-averages were
achieved by integrating over a period of more than 2900 large-eddy turnover
times. In Fig.~1 are shown the energy spectrum and mean energy flux in the
simulation.

\smallskip
{\psfig{file=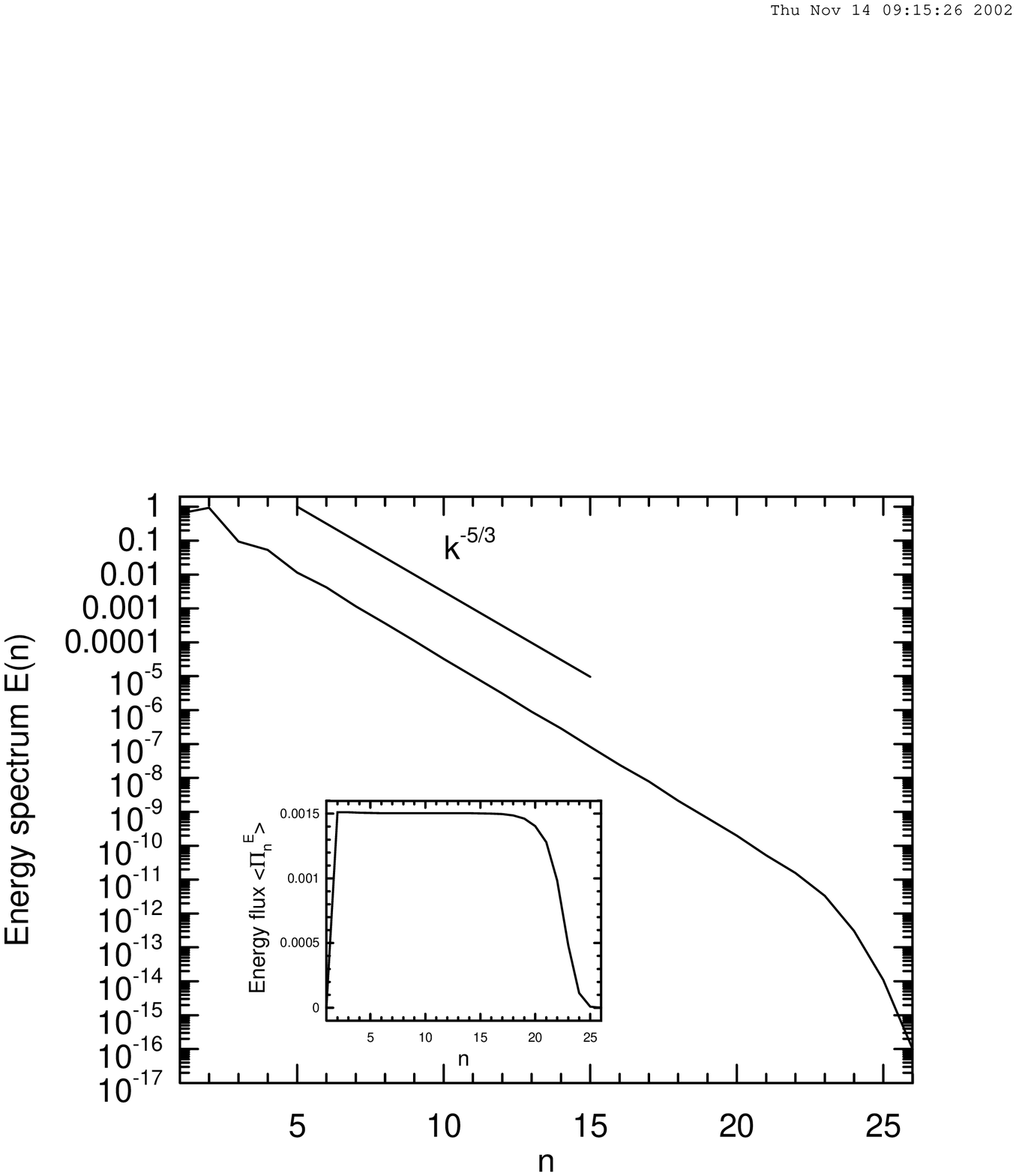,width=220pt,height=140pt}}
\noindent
{\small FIG.~1. Energy spectrum shows $k^{-5/3}$ power law from 5th shell to
23rd shell.
In the inset we show a constant mean energy flux in the inertial range. }
\smallskip

We first present evidence for the good decay of correlations of ``spins''. We
define the spin-spin
correlation functions $C_{XY}(n;m)=\langle X_n^* Y_{m}\rangle-\langle
X_n^*\rangle\langle Y_{m}\rangle$
for $X,Y=\sigma,U$. The results are shown in Figs.~2-3.
It may be seen that the correlations decay quite rapidly in $|m-n|$,
exponentially or as a
large inverse power ($\geq 7$). This is an indication that the Gibbs measure of
the hypothesis does not
correspond to a critical point with a power-law scaling. For the
``1-dimensional'' spin chain of the
shell model a phase-transition would, in any case, require a long-range
potential $\Phi$, e.g. a pair interaction decaying
by a small inverse power $\leq 2$ (e.g. see \cite{Dys-Kost}). The hypothesis of
a short-range potential
therefore rules out any such critical behavior.

\smallskip
{\psfig{file=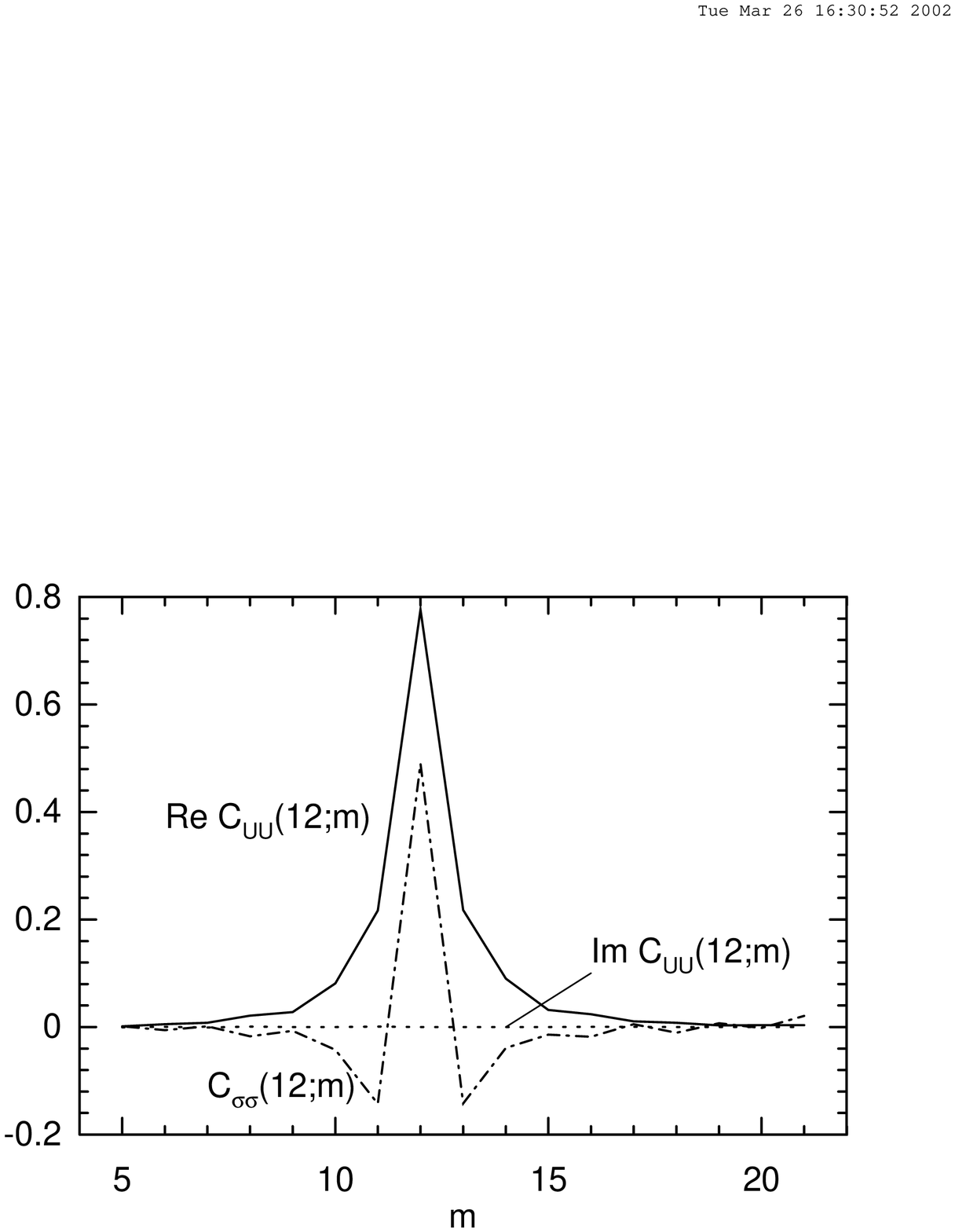,width=220pt,height=140pt}}
\noindent
{\small FIG.~2. $C_{\sigma\sigma}(n;m)$, $Re\,\,C_{UU}(n;m)$ and
$Im\,\,C_{UU}(n;m)$ for n=12.}

\smallskip
{\psfig{file=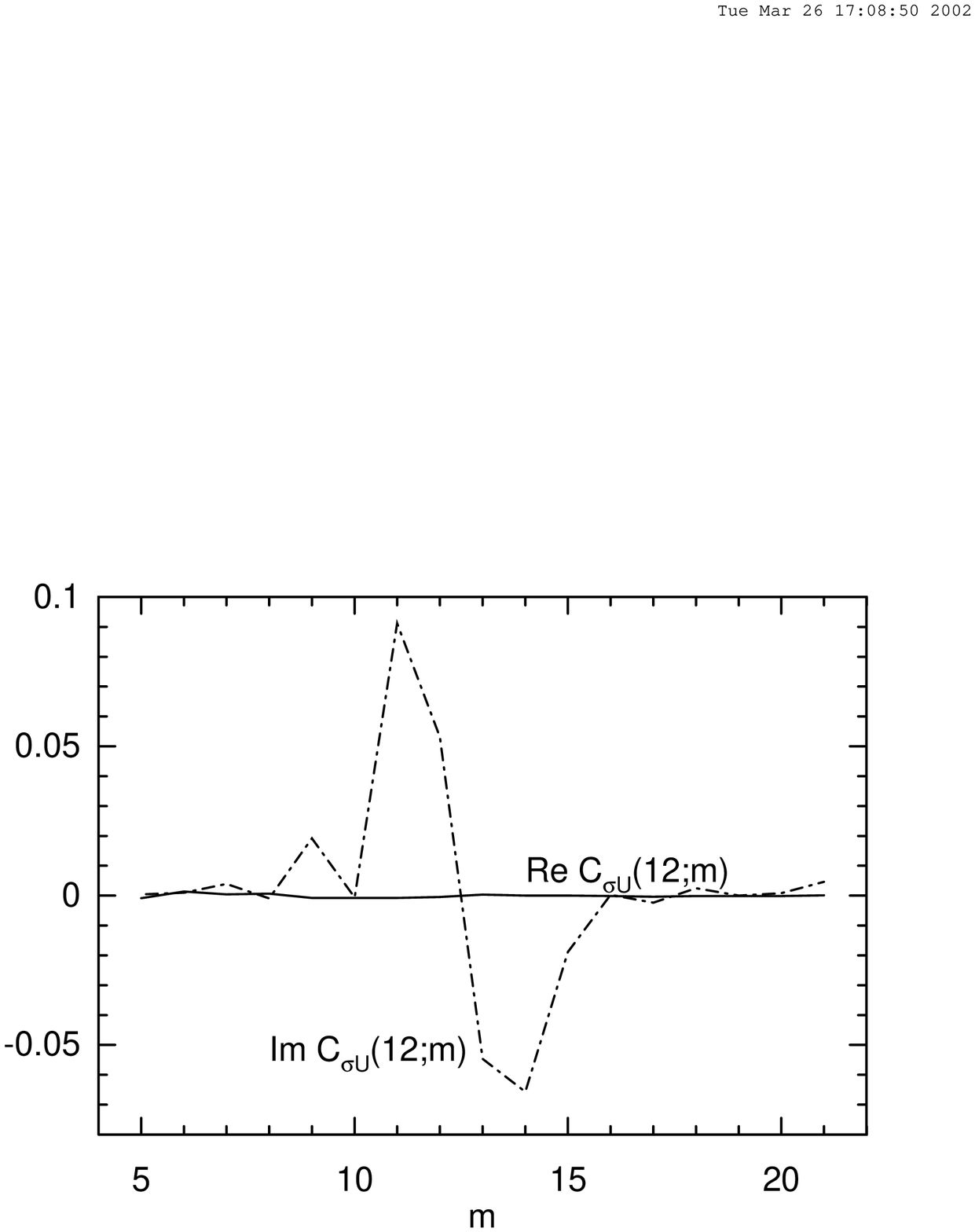,width=220pt,height=140pt}}
\noindent
{\small FIG.~3. $Re\,\,C_{\sigma U}(n;m)$ and $Im\,\,C_{\sigma U}(n;m)$ for
n=12.}
\smallskip

Further evidence against any critical behavior comes from a consideration of
the thermodynamics.
The Gibbs hypothesis implies that a ``Gibbs free-energy'' $g$ should exist for
the spin models,
defined by a suitable ``thermodynamic limit''. Thus, we may introduce
``magnetic
fields'' $p$ corresponding to the $\sigma$-spins and $h=h_x+ih_y$ corresponding
to the $U$-spins,
so that the concave free-energy is defined by
\be g(p,h)= \lim_{n\rightarrow \infty}{{-1}\over{n}}\ln Z_n(p,h) \lb{g-fun} \ee
with the ``partition function''
\be Z_n(p,h):=\langle\exp\left[\sum_{k=1}^n (p\sigma_k + {\rm Re}(h^*U_k))
                                      \right]\rangle. \lb{part-fun} \ee
The absolute structure functions are proportional to the ``partition
functions''
at $h=0$: $\langle |u_n|^p\rangle = k_n^{-p/3} Z_n(p,0)$. Thus, the existence
of a thermodynamic limit, as
implied by the Gibbs hypothesis,
yields a power-law scaling of the structure functions $\sim k_n^{-\zeta_p}$
with the anomalous exponent
$\delta\zeta_p := \zeta_p-{{p}\over{3}}$ related to the free-energy by
$g(p,0)=\delta\zeta_p \cdot
\ln\lambda.$ However, the Gibbs hypothesis implies also a power-law scaling for
the ``phase
structure-functions'' $Z_n(0,h)$. Fig.~4 shows clean power-law ranges for these
quantities,
solid evidence in favor of the Gibbs hypothesis (but not a proof, since
the thermodynamic

\smallskip
{\psfig{file= 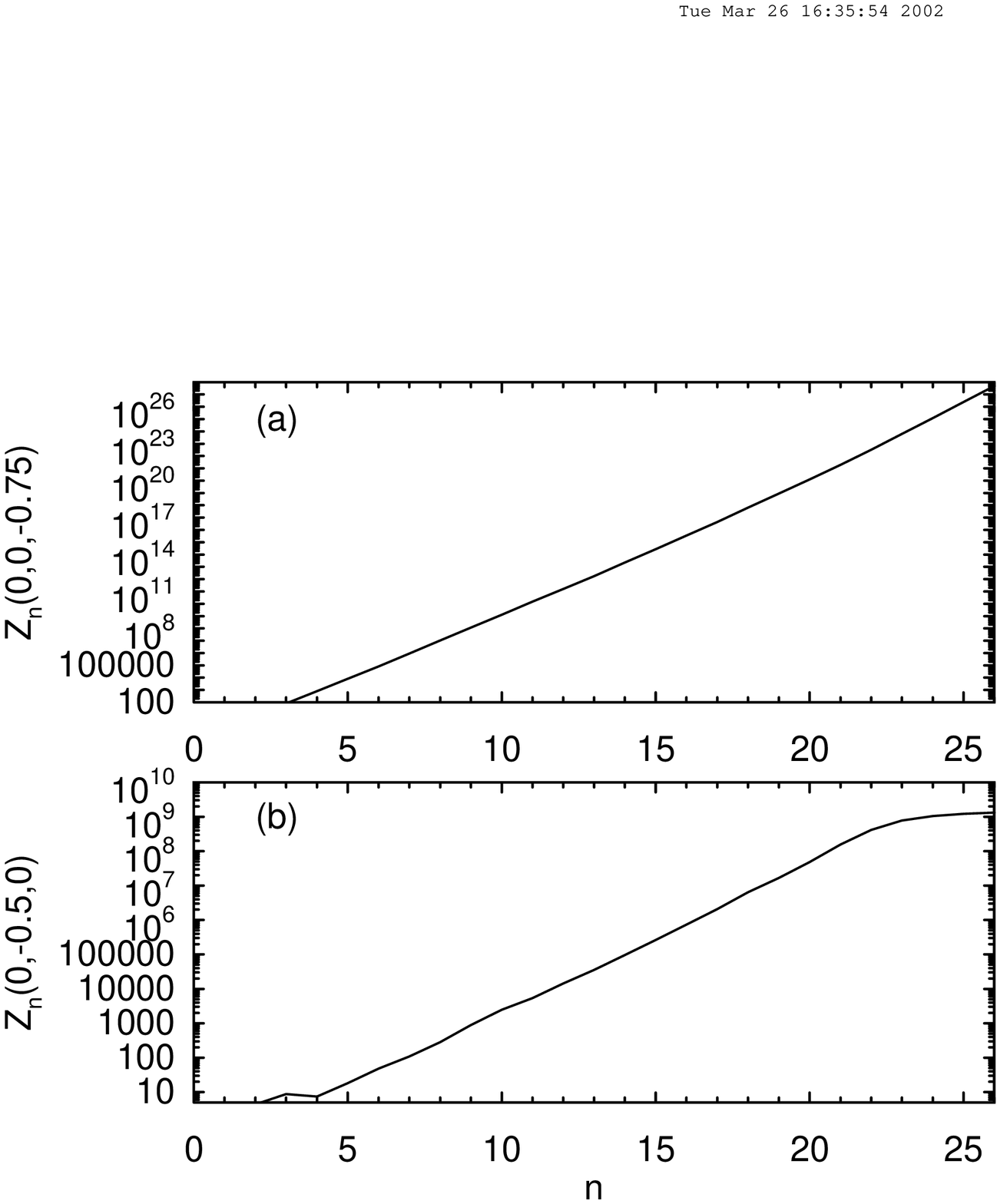,width=220pt,height=143pt}}
\noindent
{\small FIG.~4. Phase structure functions (a) $Z_n(0,0,-0.75)$; (b)
$Z_n(0,-0.5,0)$.}
\smallskip

\noindent limit could
exist even for a non-Gibbsian measure; e.g see \cite{Leb-Schon}). In Fig.~5 we
plot cross-sections
of the Gibbs free energy, $g(0,h_x,0)$ and $g(0,0,h_y)$.
\noindent These appear to be smooth functions of their arguments. There is no
evidence for any
non-analyticity that would signal appearance of a phase transition.

\smallskip
{\psfig{file=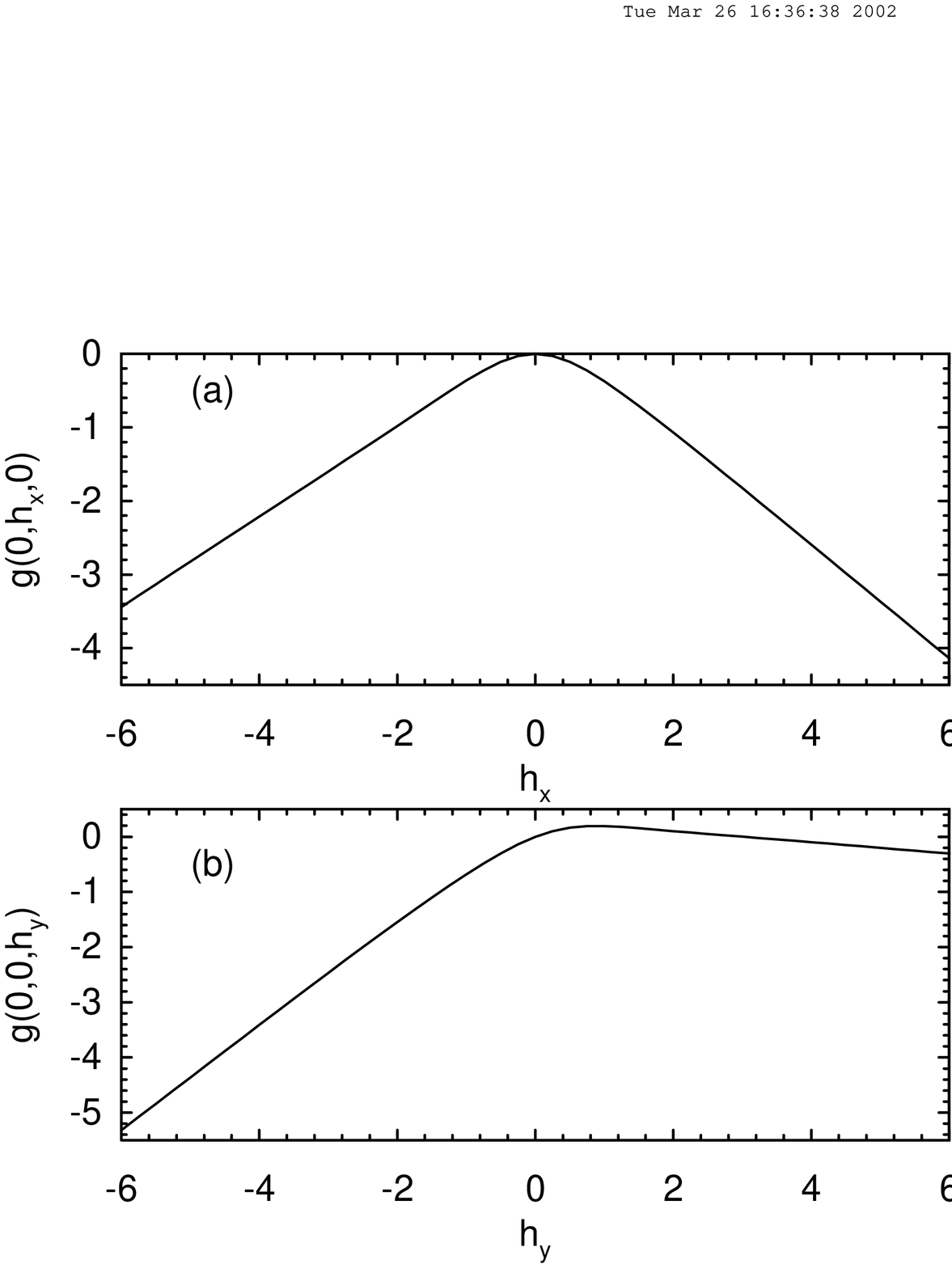,width=220pt,height=143pt}}
\noindent
\noindent
{\small FIG.~5. Gibbs free energy (a) $g(0,h_x,0)$; (b) $g(0,0,h_y)$.}
\smallskip

\indent We now study the question of ``translation-invariance'' of the
measures, by means of the single-site
spin distributions. We plot in Figs.~6 and 7 the distributions $P(\sigma_n)$
and $P(\Delta_n)$,
for different values of $n$ in the inertial range. We see that these
distributions collapse quite well,
verifying the ``translation-invariance'' assumption. The first distribution is
approximately exponential
type $P(\sigma)\approx (\alpha/2)\exp(-\alpha|\sigma|)$ with $\alpha\approx
2.00$ while the second is fit
well by $P(\Delta)\approx C\exp(-\beta\sin(\Delta))$ with $C\approx
0.1,\beta\approx 1.2$. The
``infinite-temperature'' model in \cite{BBP} does not predict well either
$P(\sigma)$ or $P(\Delta)$.
That approximation assumed a distribution $P(\Delta)$ uniform on the interval
$[-\pi,0]$ and yielded
a multiplier distribution $P(w)$ also compactly supported on a finite interval
$[w_-,w_+]$.
However, the result in Fig.~6 (see inset) implies a distribution $P(w)$ with
two power-law regimes,
$\sim w^{\alpha-1}$ for $w\ll 1$ and $\sim w^{-\alpha-1}$ for $w\gg 1.$  This
is
inconsistent not only with \cite{BBP} but with any independent spin model.
Because of the
power-law tail for large $w$, the moments which would give the anomalous
scaling exponents
for independent spins in fact diverge, $\langle w^p\rangle=+\infty$ for
$p\geq\alpha.$

\smallskip
{\psfig{file=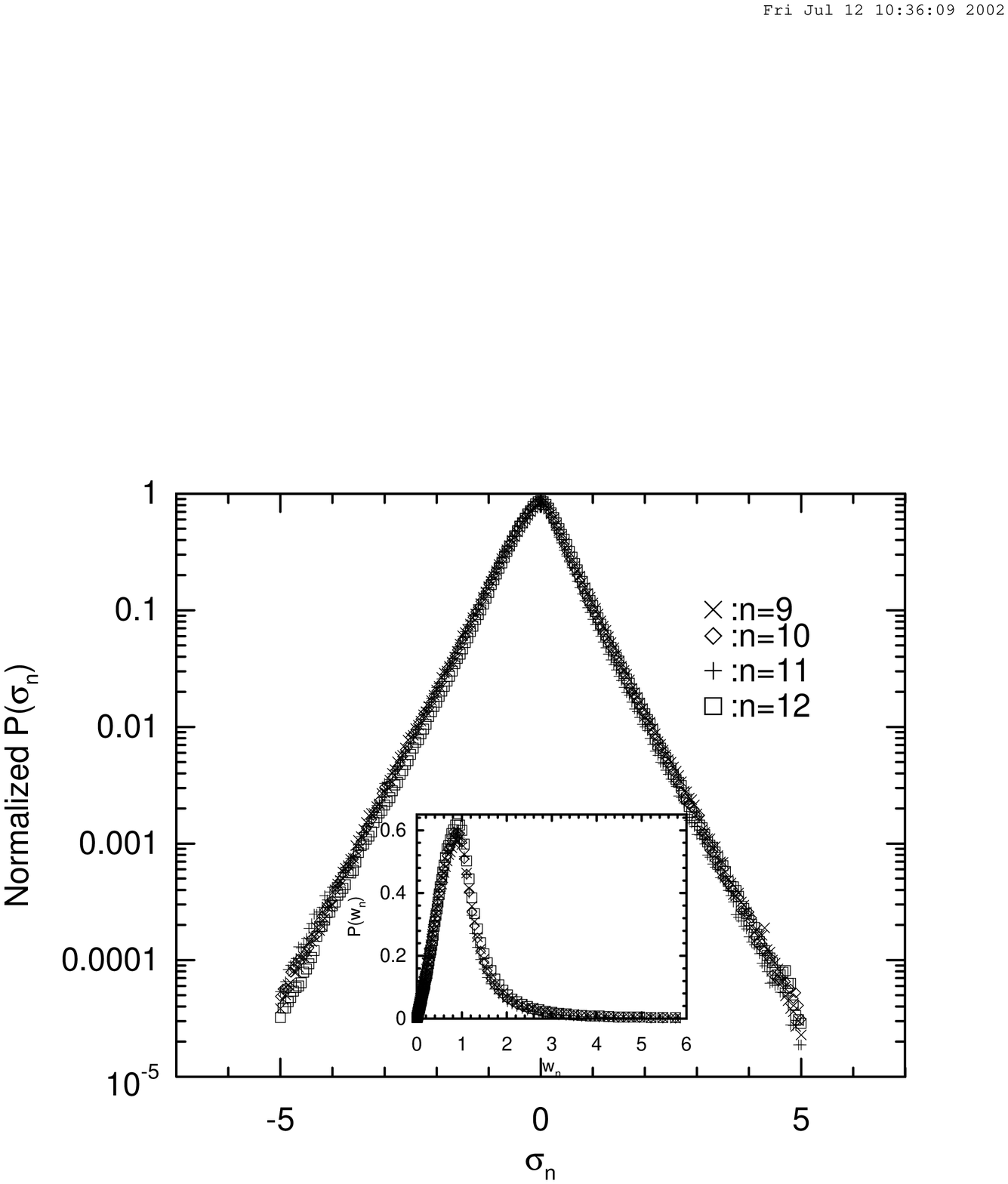,width=220pt,height=143pt}}
\noindent
\noindent
{\small FIG.~6. Distributions of  $\sigma_n$, for $n=9-12$. In the inset are
the
 distributions of $w_n$.}

\smallskip
{\psfig{file=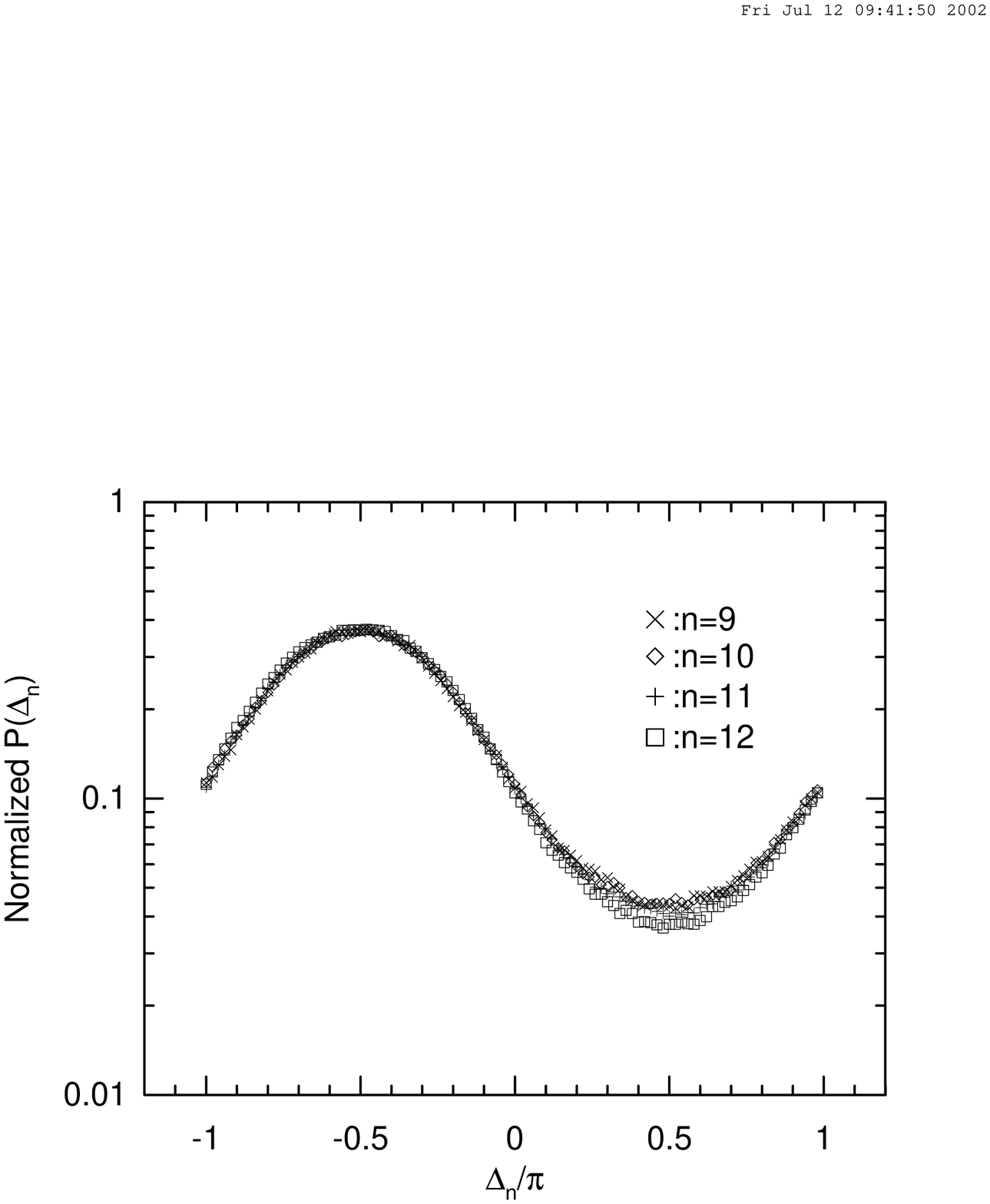,width=220pt,height=143pt}}
\noindent
\noindent
{\small FIG.~7. Distributions of $\Delta_n$, for $n=9-12$.}
\smallskip

The distributions for the $\sigma$-spins appear symmetric under spin-flip
$\sigma
\rightarrow -\sigma$ around the K41 value $\sigma=0$. In fact, the exponential
tails on the right and the
left arise symmetrically from the same events in SABRA, which we call
``defects''. These are events
in which the amplitude in a single shell, say, the $n$th, drops to a very low
value, $\rho_n\ll 1$.
Intermittent bursts in the shell models are generally preceded by such events
\cite{Gil-Domb,Ok-Jens}, but defects do not need to appear in association with
bursts. Since $\sigma_n =
\ln(\rho_n/\rho_{n-1}),$ the negative tail where $\sigma_n\ll -1$ comes from
realizations with a ``defect''
in the $n$th shell, while the positive tail where $\sigma_n\gg 1$ comes from
realizations with ``defects'' in
the $(n-1)$st shell. The statistical distribution of such ``defects'' in a
given shell $n$ can be inferred
from the constancy of $P(u_n)$ at $u_n=0$. By a change to polar coordinates
$(\rho_n,\theta_n)$, one
finds $P(\rho_n)\approx (const.)\rho_n$ for $\rho_n\ll 1.$ For events in the
negative tail, $\rho_n\ll 1$
while $\rho_{n-1}\approx 1.$ In that case, $\sigma_n\approx \ln\rho_n$ and
$\rho_n \approx e^{\sigma_n}$.
Therefore, by change of variables, $P(\sigma_n) = \left.
\left|{{d(\rho_n)}\over{d\sigma_n}}\right|
\cdot P(\rho_n)\right|_{\rho_n\approx e^{\sigma_n}}\approx (const.)
e^{2\sigma_n}$ for $\sigma_n\ll -1$.
An identical argument for the positive tail using $\sigma_n\approx
-\ln(\rho_{n-1})$ and $\rho_{n-1}
\approx e^{-\sigma_n}$ gives $P(\sigma_n) \approx (const.) e^{-2\sigma_n}$ for
$\sigma_n\gg 1$. This argument assumes only that the amplitudes
$\rho_n,\rho_{n-1}$ in the ratio are not strongly correlated, which
could suppress the long tails.
The distribution $P(\Delta)$ also exhibits a symmetry $\Delta\rightarrow
\pi-\Delta$ around the K41
solution with $\Delta=-\pi/2$. This symmetry can be expressed as $U\rightarrow
-U^*$
and is seen as well in the vanishing of ${\rm Im}\,C_{UU}$ and ${\rm
Re}\,C_{\sigma U}$
in Figs.~2-3 and in the near symmetry of $g(0,h_x,0)$ in Fig.~5.
In fact, this is an exact symmetry of the dynamics, broken only by our forcing.
If $u_n$ is a
solution of SABRA with a force $f_n$, then $-u_n^*$ is a solution with force
$-f_n^*$ and under this
transformation $U_n\rightarrow -U_n^*$ for all $n$. This symmetry should be
restored for
large $n$, similar to restoration of isotropy in 3D.

The distribution $P(\sigma)$ cannot be exactly symmetrical under the spin-flip
$\sigma\rightarrow -\sigma$.
There must be a non-vanishing mean $\langle\sigma\rangle$ or a
``magnetization'', due to the fact that
Kolmogorov 1941 mean-field scaling of $p$th-order structure functions is not
exact at $p=0$ \cite{SVBSCC}.
Indeed, by the Gibbs hypothesis, the magnetization can be obtained from the
thermodynamic formula
$\langle\sigma\rangle= -\left.{{\partial g}\over{\partial p}}\right|_{p,h=0}.$
In Fig.~8 we plot
$\langle\sigma_n\rangle$ vs. $n$ for the two simulations of SABRA

\smallskip
{\psfig{file=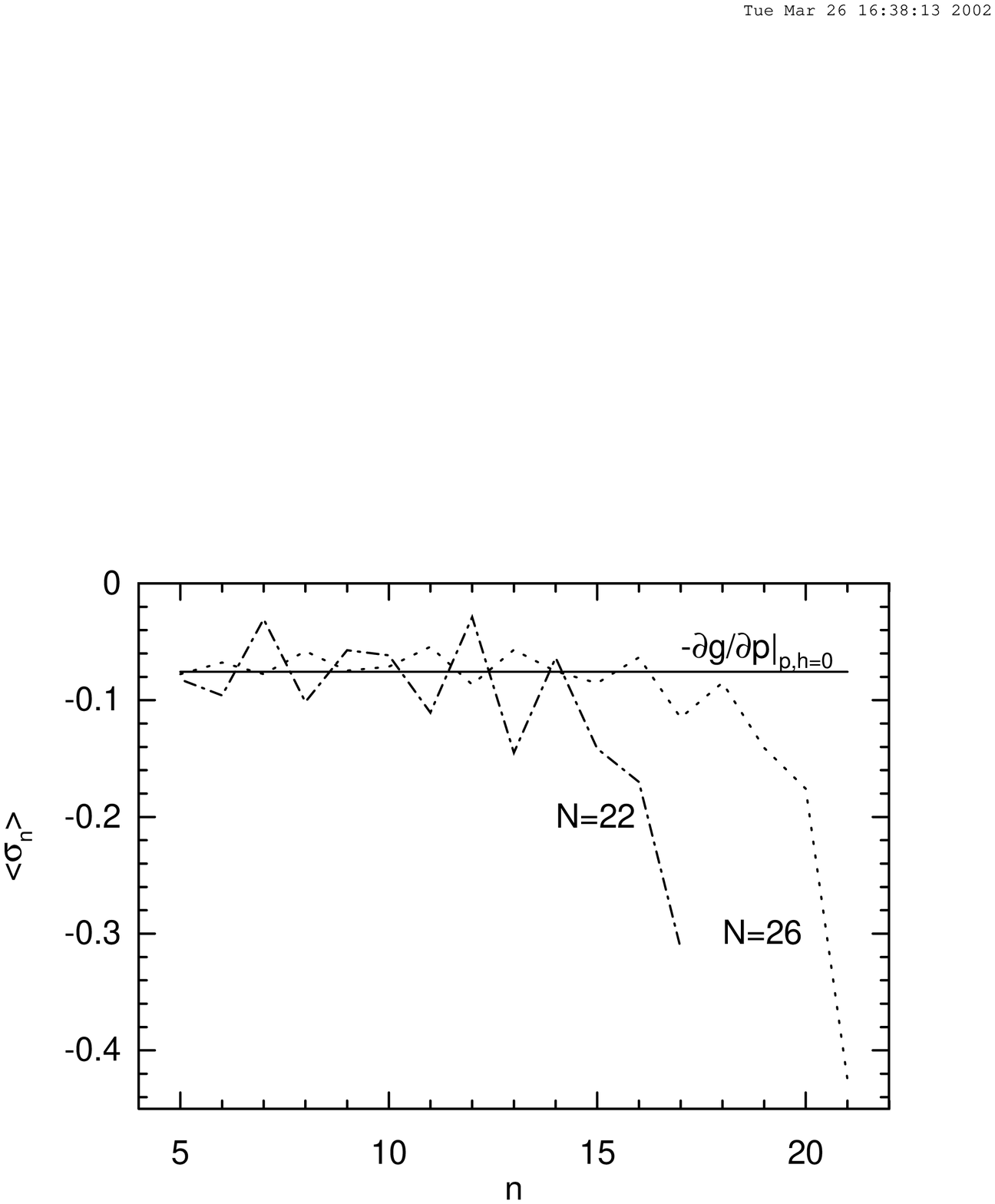,width=220pt,height=143pt}}
\noindent
{\small FIG.~8.  $\langle\sigma_n\rangle$ vs. $n$ with $-\left.\partial
g/\partial p\right|_{p,h=0}$.}

{\psfig{file=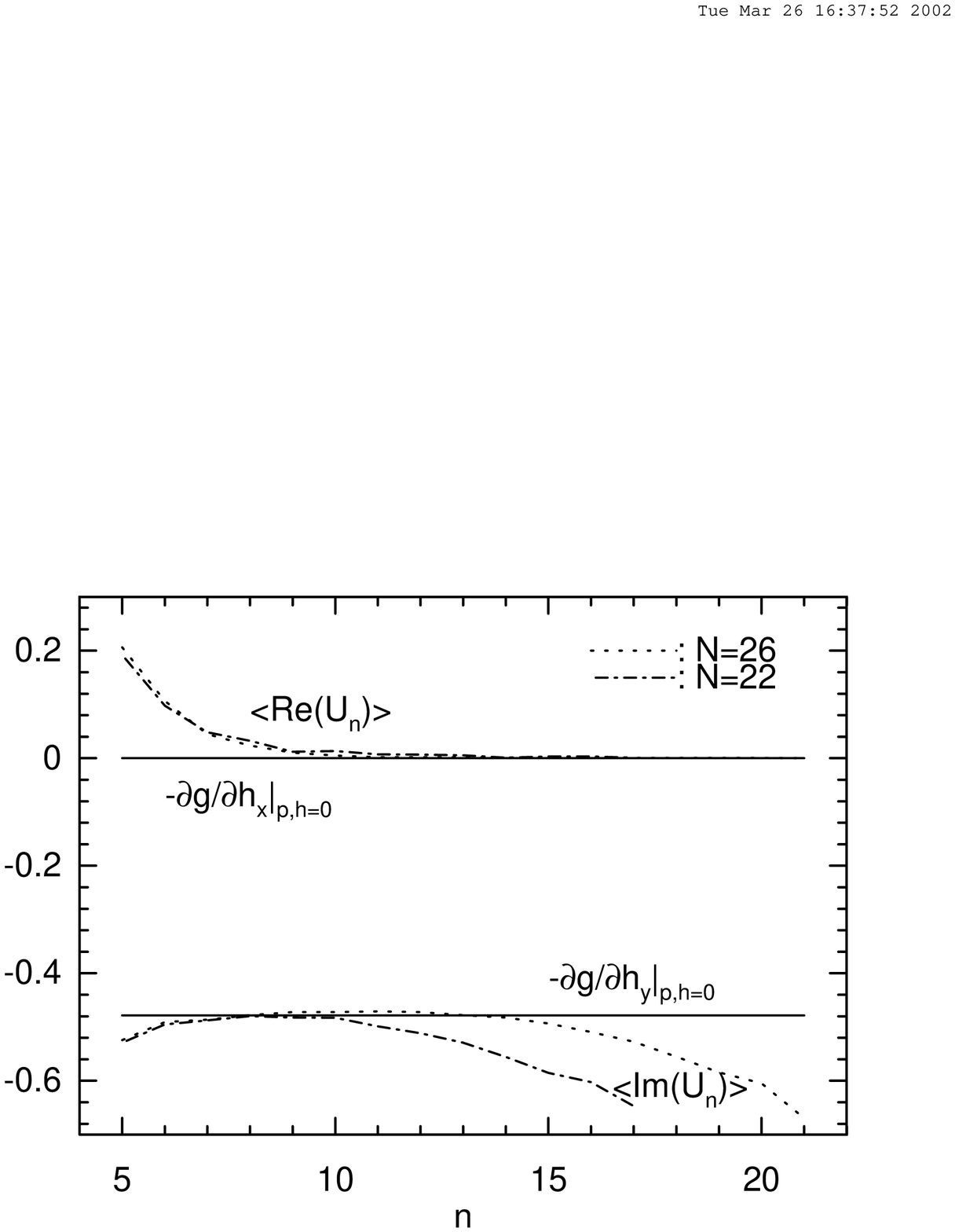,width=220pt,height=143pt}}
\noindent
{\small FIG.~9.  $\langle Re(U_n)\rangle$ vs. $n$ with $-\left.\partial
g/\partial h_x\right|_{p,h=0}$
and of $\langle Im(U_n)\rangle$ vs. $n$ with $-\left.\partial g/\partial
h_y\right|_{p,h=0}$. }
\smallskip

\noindent with $N=22$ and $N=26$. We see that
there is a slight breaking of translation symmetry. This is likely due to
finite-size effects and becomes
much smaller for the $N=26$ shell simulation than for $N=22$. In the same
figure is plotted a straight
horizontal line for the prediction $\langle\sigma\rangle\approx -0.0757$
obtained from the derivative
of $g$. We see that the agreement is quite satisfactory. It is also apparent
from Fig.~7 that a non-zero
expectation $\langle U\rangle$ occurs, non-invariant under conjugation
$U\rightarrow U^*$.
In fact, a negative value $\langle\sin(\Delta)\rangle<0$ is required in SABRA
for a forward cascade
of energy \cite{BBP,LPPPV}. We plot in Fig.~9 the ``magnetizations'' $\langle
U_n\rangle$ vs. wavenumber
$n$. Again we see a finite-size breaking of translation-symmetry, which lessens
going from $N=22$ to
$N=26.$ The numerical values of these ``magnetizations'' are also predicted by
derivatives of the Gibbs
free energy, giving a large expectation $\langle\sin\Delta\rangle \approx
-0.489$ but a small value
$\langle \cos\Delta\rangle = 0.00814,$ consistent with approximate symmetry
$U\rightarrow -U^*$.
These results are plotted in Figs.~8 and 9 as horizontal lines and obviously
give satisfactory agreement
with the direct measurements of magnetizations. The results indicate that there
is ``spin-ordering''
in the turbulent systems, which breaks discrete spin-flip symmetries.
Peierls-type arguments \cite{Peierls}, including rigorous versions such
as Pirogov-Sinai theory \cite{PS}, indicate
that discrete symmetries cannot be spontaneously broken in
``1-dimensional lattice'' systems such as the shell model, if the interaction
potential is short-ranged.
A 1-dimensional Gibbs distribution with non-zero magnetization for a symmetric,
short-ranged potential
is unstable to formation of domain walls. We expect the spin order here to be
not ``ferromagnetic'' but
``paramagnetic'', arising from explicit symmetry-breaking terms in the
potential $\Phi$ of the Gibbs measure.

\section{Discussion and Conclusions}

In this work, we have shown that Kolmogorov multipliers and dynamical phases
in the shell model must be correlated from shell-to-shell. We have also
presented evidence that these ``spin'' variables, while not independent,
are distributed according to a translation-invariant Gibbs measure.
Of course, this has certainly not been proved in this work. There are
non-Gibbsian measures on one-dimensional lattices, such as that of Schonmann
\cite{Schon}, which have both exponential decay of correlations and a
thermodynamic
limit of the Gibbs free energy (which, however, is there non-analytic).
On the other hand, that type of example has been shown to be Gibbs also
in a somewhat generalized sense \cite{DobShlos}. The hypothesis that the
distribution of suitably defined ``spins'' in the shell model is Gibbsian
with a summable potential has testable consequences, some of which we have
verified in this work.

If the hypothesis is true, then it is possible to recover the potentials
from the finite-shell marginal distributions. Indeed, if $P_{n,...,n+N-1}
(\xi_n,...,\xi_{n+N-1})$ is the probability density of the ``spins'' at shells
$n,...,n+N-1$ in the inertial range, then
\be \ln P_{n,...,n+N-1}(\xi_n,...,\xi_{n+N-1})
                   = -\sum_{k=0}^{N-1} \Phi_{n+k}(\xi) + o(N), \lb{Hamilt} \ee
whenever the potential is absolutely summable. See Proposition 2.46 in
\cite{EFS}. More directly, the potentials may be derived from conditional
probabilities of the spins via the M\"{o}bius inversion formula \cite{KSK}.
For example, if the ``spins'' are distributed by a Markov chain or
nearest-neighbor Gibbs measure, then, up to constants, the 1-body
interaction is
\be \Phi^{(1)}_n(\xi_n) = -\ln T_{n|n-1}(\xi_n|\xi^*)
                          -\ln T_{n+1|n}(\xi^*|\xi_n)   \lb{1-body} \ee
and the 2-body interaction is
\begin{eqnarray}
\Phi^{(2)}_{n,n-1}(\xi_n,\xi_{n-1}) & = &  -\ln T_{n|n-1}(\xi_n|\xi_{n-1}) \cr
\, & & \,\,\,\,\,\,\,\,\,\,\,\,\,\,\,\,
       +\ln T_{n|n-1}(\xi_n|\xi^*)+\ln T_{n|n-1}(\xi^*|\xi_{n-1}).
   \lb{2-body}
\end{eqnarray}
Here $T_{n|n-1}(\xi_n|\xi_{n-1})$ is the transition probability of the
Markov chain and $\xi^*$ is a constant reference value of the spin.
As this example makes clear, the potentials of the Gibbs measure can
be recovered, in principle, from conditional probabilities obtained
in numerical simulations of the dynamics.

If true, it is intellectually interesting that probability measures
arising from turbulent dynamics may be Gibbsian, in a suitable spin
representation. However, more importantly, it gives one some new tools
that one may apply to the turbulence problem. For example, it offers
a new route to calculate the scaling exponents as a ``free energy''
$g(p,h)$ of a one-dimensional spin system. Furthermore, formulas
for the conditional probabilities of small-scale modes given the
large-scale ones provided by the Gibbs hypothesis may be very useful
in carrying out large-eddy simulations of turbulence \cite{MK}.
Some preliminary tests of these ideas have already been carried
out for the shell models and will be reported elsewhere \cite{ECC}.

Similar results as discussed here hold for Navier-Stokes dynamics
\cite{CCE}, using the velocity increments advocated by Kolmogorov
\cite{K62}. Multipliers and spins may also be defined by a
representation of the Navier-Stokes equations in terms of orthogonal
wavelet bases \cite{Meneveau}. In that formulation, the dynamics
resembles the shell model on the dyadic Cayley tree \cite{BBTT}.
Generalizing \cite{BBP}, the invariant measure of the latter should
be a Gibbs measure on a Bethe lattice (\cite{Baxter}, Ch.4.) Whereas
standard shell models corespond to ``1-dimensional'' spin systems,
the spin systems on the Bethe lattice are effectively ``infinite-dimensional''.
Nevertheless, the statistics in the shell model and in Navier-Stokes
should be qualitatively similar, as present evidence suggests that
the Gibbs distributions for both are in the high-temperature,
unique-phase regime.

\setcounter{section}{0}
\renewcommand{\thesection}{\Alph{section}}

\section{Appendices}

\subsection{Stabilization of the K41 Solution}

We show in this first appendix that the K41 solution would be stable
if multipliers were independent for distinct shells (or, in fact, even if
they were independent for shells a finite distance apart). We show that
this follows from a set of exact dynamical constraints on the multiplier
variables $\xi_n=(w_n,\Delta_n)$.

For this purpose, we must transform the equations of motion into those
variables. In terms of $\rho_n,\theta_n$, the SABRA dynamics becomes
\begin{eqnarray}
{{\dot{\rho}_n}\over{\rho_n}} & = & {{k_n^{2/3}}\over{\lambda}}
   \left[ a {{\rho_{n+2}\rho_{n+1}\over{\rho_n}}} \sin\Delta_{n+2}
         + b {{\rho_{n+1}\rho_{n-1}\over{\rho_n}}} \sin\Delta_{n+1}
         +c {{\rho_{n-1}\rho_{n-2}\over{\rho_n}}} \sin\Delta_n\right] \cr
    \, & &
\,\,\,\,\,\,\,\,\,\,\,\,\,\,\,\,\,\,\,
\,\,\,\,\,\,\,\,\,\,\,\,\,\,\,\,\,\,\,\,\,
+ {{k_n^{1/3}{\rm Re}(f_ne^{-i\theta_n})\over{\rho_n}}} -\nu k_n^2
\lb{rho-eq}
\end{eqnarray}
and
\begin{eqnarray}
\dot{\theta}_n & = & {{k_n^{2/3}}\over{\lambda}}
   \left[ a {{\rho_{n+2}\rho_{n+1}\over{\rho_n}}} \cos\Delta_{n+2}
         + b {{\rho_{n+1}\rho_{n-1}\over{\rho_n}}} \cos\Delta_{n+1}
         -c {{\rho_{n-1}\rho_{n-2}\over{\rho_n}}} \cos\Delta_n\right] \cr
    \, & &
\,\,\,\,\,\,\,\,\,\,\,\,\,\,\,\,\,\,\,\,\,\,\,\,\,\,\,\,\,\,\,\,\,\,\,\,
\,\,\,\,\,\,\,\,\,\,\,\,\,\,\,\,\,\,\,\,\,\,\,\,\,\,\,\,\,\,\,\,\,\,\,\,
         + {{k_n^{1/3}{\rm Im}(f_ne^{-i\theta_n})\over{\rho_n}}}.
\lb{theta-eq}
\end{eqnarray}
Going over to the scale-local variables $w_n,\Delta_n$, this becomes
$$ \dot{w}_n = w_n [U_n(w,\Delta) - U_{n-1}(w,\Delta)]:= W_n(w,\Delta) $$
and
$$ \dot{\Delta}_n = -V_n(w,\Delta) + V_{n-1}(w,\Delta) + V_{n-2}(w,\Delta)
                  := Z_n(w,\Delta) \lb{Delta-eq} $$
where $U_n,V_n$ are the righthand sides of (\ref{rho-eq}),(\ref{theta-eq}),
respectively, expressed in terms of $w_n,\Delta_n$. Thus, considering just
the inertial-range part of the dynamics,
\begin{eqnarray}
U_n(w,\Delta) & = & {{k_n^{2/3}}\over{\lambda}} \cdot
\left[ a \cdot w_{n+2}\sin\Delta_{n+2}\cdot w_{n+1}^2 w_n w_{n-1} \right. \cr
\, & & \,\,\,\,\,\,\,\,
       + \left. b \cdot w_{n+1}\sin\Delta_{n+1} \cdot w_{n-1}  +c w_n^{-1}
\sin\Delta_n\right]
        \prod_{k=1}^{n-2} w_k
\end{eqnarray}
and
\begin {eqnarray}
V_n(w,\Delta) & = & {{k_n^{2/3}}\over{\lambda}} \cdot
\left[ a \cdot w_{n+2}\cos\Delta_{n+2}\cdot w_{n+1}^2 w_n w_{n-1} \right. \cr
\, & &  \,\,\,\,\,\,\,\,
        + \left. b \cdot w_{n+1}\cos\Delta_{n+1} \cdot w_{n-1}-c
w_n^{-1}\cos\Delta_n\right]
        \prod_{k=1}^{n-2} w_k
\end{eqnarray}
In terms of these variables, the dynamics appears highly nonlocal in scale,
because of the product $\prod_{k=1}^{n-2} w_k$.

If $P_n(\xi_n)$ is the distribution of $\xi_n=(w_n,\Delta_n)$, then it is
straightforward to show that
\be \partial_t P_n(\xi_n) =
-{{\partial}\over{\partial w_n}}\left[ \langle W_n|\xi_n\rangle
P_n(\xi_n)\right]
-{{\partial}\over{\partial \Delta_n}}\left[ \langle Z_n|\xi_n\rangle
P_n(\xi_n)\right],
\lb{Pn-eq} \ee
where $\langle W_n|\xi_n\rangle ,\langle Z_n|\xi_n\rangle$ are conditional
averages
for fixed $\xi_n$. These equations are exact, but not closed in terms of $P_n$.
However, if we assume that the statistics of the model are given by a product
measure $\prod_n P_n(\xi_n)$---and with {\it only} that assumption---then we
obtain
closed equations for the $P_n$'s. In the inertial range, these equations are
of the form (\ref{Pn-eq}) with
$$  \langle W_n|\xi_n\rangle
= {{k_n^{2/3}}\over{\lambda}} \prod_{k=1}^{n-4} \langle w_k \rangle
\left[ (A_n w_n^3 - B_n w_n^2 + C_n w_n) + (D_n w_n^2 - E_n)\sin\Delta_n\right]
$$
$$ \langle Z_n|\xi_n\rangle
= {{k_n^{2/3}}\over{\lambda}} \prod_{k=1}^{n-4} \langle w_k \rangle
\left[ (F_n w_n^2 + G_n w_n + H_n) - (K_n w_n^{-1} + L_n
w_n)\cos\Delta_n\right] $$
where
$$ A_n = -{{a}\over{\lambda^{2/3}}} \langle w_{n+1}S_{n+1}\rangle
          \langle w_{n-1}\rangle\langle w_{n-2}\rangle\langle w_{n-3}\rangle
$$
$$ B_n = - a \langle w_{n+2}S_{n+2}\rangle \langle w_{n+1}^2\rangle
          \langle w_{n-1}\rangle\langle w_{n-2}\rangle\langle w_{n-3}\rangle
$$
$$ C_n = b \langle w_{n+1}S_{n+1}\rangle\langle w_{n-1}\rangle\langle
w_{n-2}\rangle\langle w_{n-3}\rangle
         - {{c}\over{\lambda^{2/3}}}
\langle{{S_{n-1}}\over{w_{n-1}}}\rangle\langle w_{n-3}\rangle  $$
$$ D_n = -{{b}\over{\lambda^{2/3}}} \langle w_{n-2}\rangle\langle
w_{n-3}\rangle  $$
$$ E_n = - c  \langle w_{n-2}\rangle\langle w_{n-3}\rangle  $$
$$ F_n = {{a}\over{\lambda^{2/3}}} \langle w_{n+1}C_{n+1}\rangle
          \langle w_{n-1}\rangle\langle w_{n-2}\rangle\langle w_{n-3}\rangle
$$
$$ G_n = - a \langle w_{n+2}C_{n+2}\rangle \langle w_{n+1}^2\rangle
          \langle w_{n-1}\rangle\langle w_{n-2}\rangle\langle w_{n-3}\rangle
$$
$$
H_n = -b \langle w_{n+1}C_{n+1}\rangle\langle w_{n-1}\rangle\langle
w_{n-2}\rangle\langle w_{n-3}\rangle
         - {{c}\over{\lambda^{2/3}}}
\langle{{C_{n-1}}\over{w_{n-1}}}\rangle\langle w_{n-3}\rangle $$
$$ \,\,\,\,\,\,\,\,\,\,\,\,\,\,\,\,\,\,\,\,\,\,\,\,\,\,\,\,\,\,\,\,\,\,
           + {{b}\over{\lambda^{4/3}}} \langle w_{n-1}C_{n-1}\rangle \langle
w_{n-3}\rangle
           - {{c}\over{\lambda^{4/3}}} \langle{{C_{n-2}}\over{w_{n-2}}}\rangle
$$
$$ K_n = -c \langle w_{n-2}\rangle\langle w_{n-3}\rangle  $$
$$ L_n = -{{b}\over{\lambda^{2/3}}} \langle w_{n-2}\rangle\langle
w_{n-3}\rangle
         -{{a}\over{\lambda^{4/3}}} \langle w_{n-1}^2\rangle \langle
w_{n-2}\rangle\langle w_{n-3}\rangle.  $$
In these expressions $S_n=\sin\Delta_n,\,\,C_n=\cos\Delta_n$ for all $n$. Note
that the
resulting closed equations for the $P_n$'s are nonlinear integro-partial
differential equations,
since the averages in the above expressions are over the $P_n$'s themselves.

If we assume further that all the ``spins'' $\xi_n$ are identically
distributed, i.e. $P_n=P$
for all $n$, then the above equations for $P_n$ for each $n$ reduce to the same
equation for $P$,
after a change in the time-scale by a factor
${{k_n^{2/3}}\over{\lambda}}\langle w\rangle^{n-4}$:
\be \partial_t P(\xi)
= -{{\partial}\over{\partial w}}\left[ \overline{W}(\xi) P(\xi)\right]
-{{\partial}\over{\partial \Delta}}\left[ \overline{Z}(\xi) P(\xi)\right],
\lb{P-eq} \ee
with now
$$  \overline{W}(\xi) =  (A w^3 - B w^2 + C w) + (D w^2 - E)\sin\Delta $$
$$  \overline{Z}(\xi) =  (F w^2 + G w + H) - (K w^{-1} + L w)\cos\Delta $$
where
$$ A = -{{a}\over{\lambda^{2/3}}} \langle w S\rangle \langle w\rangle^3  $$
$$ B = - a \langle wS\rangle \langle w^2\rangle \langle w\rangle^3  $$
$$ C = b \langle wS\rangle \langle w\rangle^3 - {{c}\over{\lambda^{2/3}}}
\langle{{S}\over{w}}\rangle\langle w\rangle  $$
$$ D = -{{b}\over{\lambda^{2/3}}} \langle w\rangle^2  $$
$$ E = - c  \langle w\rangle^2  $$
$$ F = {{a}\over{\lambda^{2/3}}} \langle wC\rangle \langle w\rangle^3  $$
$$ G = - a \langle wC\rangle \langle w^2\rangle \langle w\rangle^3 $$
$$ H = -b \langle wC\rangle\langle w\rangle^3
         - {{c}\over{\lambda^{2/3}}} \langle{{C}\over{w}}\rangle\langle
w\rangle
           + {{b}\over{\lambda^{4/3}}} \langle wC\rangle \langle w\rangle
           - {{c}\over{\lambda^{4/3}}} \langle{{C}\over{w}}\rangle $$
$$ K = -c \langle w\rangle^2 $$
$$ L = -{{b}\over{\lambda^{2/3}}} \langle w\rangle^2
       -{{a}\over{\lambda^{4/3}}}\langle w^2\rangle\langle w\rangle^2.  $$
Using $a+b+c=0$, it is easy to verify that
$P_\pm(\xi)=\delta(w-1)\delta(\Delta\mp {{\pi}\over{2}})$
are exact time-independent solutions of (\ref{P-eq}). The presence of these
solutions is due to the
well-known existence of exact, steady-state ``K41'' solutions of the SABRA
model, of the
form $u_n^\pm = \pm i A k_n^{-1/3}$ for any choice of the real constant $A>0$.
The solution
$u_n^+$ has a backward energy transfer to low wavenumbers, while $u_n^-$ has
forward transfer
to high wavenumbers.

It is interesting that $P_-(\xi)$ is linearly stable under the dynamics
(\ref{P-eq}),
while $P_+(\xi)$ is unstable. In fact, the linearization of (\ref{P-eq}) is
\begin{eqnarray}
\partial_t \delta P(\xi) & = & -{{\partial}\over{\partial w}}\left[
\overline{W}_\pm(\xi) \delta P(\xi)\right]
                      -{{\partial}\over{\partial \Delta}}\left[
\overline{Z}_\pm(\xi) \delta P(\xi)\right] \cr
\,       &  & -{{\partial}\over{\partial w}}\left[ \delta \overline{W}(\xi)
P_\pm (\xi)\right]
                      -{{\partial}\over{\partial \Delta}}\left[ \delta
\overline{Z}_\pm(\xi) P_\pm (\xi)\right]
\lb{linP-eq}
\end{eqnarray}
with
\be \overline{W}_\pm (\xi)
=  (A_\pm w^3 - B_\pm w^2 + C_\pm w) + (D_\pm w^2 - E_\pm)\sin\Delta
\lb{Wpm-eq} \ee
\be \overline{Z}_\pm (\xi)
=  - (K_\pm w^{-1} + L_\pm w)\cos\Delta \lb{Zpm-eq} \ee
where
$$ A_\pm = \mp {{a}\over{\lambda^{2/3}}}, B_\pm  = \mp a, C_\pm = \pm \left(b -
{{c}\over{\lambda^{2/3}}} \right),
   D_\pm  = -{{b}\over{\lambda^{2/3}}}, E_\pm = - c, $$
$$ F_\pm = G_\pm = H_\pm =0, K_\pm  = -c, L_\pm  =
-\left({{b}\over{\lambda^{2/3}}} +{{a}\over{\lambda^{4/3}}}\right).  $$
and $\delta \overline{W}(\xi),\delta \overline{Z}(\xi)$ have the same form as
$\overline{W}(\xi),\overline{Z}(\xi)$
but with coefficients $\delta A,\delta B,...,$ etc. that can be obtained by
linearizing the corresponding
coefficients $A,B,...$, etc. Now the essential fact is that $\xi_\pm=(1,\pm
{{\pi}\over{2}})$ is an unstable/stable
fixed point of the dynamical system $(\dot{w},\dot{\Delta})=(\overline{W}_\pm
(\xi),\overline{Z}_\pm (\xi))$,
for $\pm$ respectively. This is easily verified directly from the equations
(\ref{Wpm-eq}),(\ref{Zpm-eq}).
For example, the linearization about those fixed points is
$$ \left. \left( \begin{array}{cc}
                {{\partial\overline{Z_-}}\over{\partial w}} &
{{\partial\overline{Z_-}}\over{\partial \Delta}} \cr
                {{\partial\overline{W_-}}\over{\partial w}} &
{{\partial\overline{W_-}}\over{\partial \Delta}}
                 \end{array} \right) \right|_{\xi=\xi_+}
         =  \left( \begin{array}{cc}
                      \left(1-{{1}\over{\lambda^{2/3}}}\right)(a-c) & 0 \cr
                       0 & \left(1-{{1}\over{\lambda^{4/3}}}\right)a +
\left(1-{{1}\over{\lambda^{2/3}}}\right)b
                      \end{array}\right). $$
$$ \left. \left( \begin{array}{cc}
                {{\partial\overline{Z_+}}\over{\partial w}} &
{{\partial\overline{Z_+}}\over{\partial \Delta}} \cr
                {{\partial\overline{W_+}}\over{\partial w}} &
{{\partial\overline{W_+}}\over{\partial \Delta}}
                 \end{array} \right) \right|_{\xi=\xi_-}
         = - \left( \begin{array}{cc}
                      (2a+b) + {{2a+5b}\over{\lambda^{2/3}}} & 0 \cr
                       0 & \left(1-{{1}\over{\lambda^{4/3}}}\right)a +
\left(1-{{1}\over{\lambda^{2/3}}}\right)b
                      \end{array}\right). $$
When $a>0$ and $b,c<0$, then we see that both eigenvalues are positive for the
$+$ fixed point, and
the second eigenvalue is negative for the $-$ fixed point. The first eigenvalue
of the $-$ fixed point
is also negative when
$$ 2a+5b>-(2a+b)\lambda^{2/3}, \lb{cond} $$
which imposes a condition on the coefficients. For example, with the common
parameterization $a=1,b=-\en,c=\en-1$,
a region in the $(\en,\lambda)$-plane is selected specified by
$2-5\en>(\en-2)\lambda^{2/3}$. Along the curve
$\lambda=1/(1-\en)$ for which the second invariant is helicity-like, the
condition is that $\lambda$ should lie
in an interval $(1,\lambda_*)$ with $\lambda_*\approx 2.467$. In particular,
the standard case $\lambda=2,a=1,
b=c=-1/2$ which we simulated in this paper lies in this region. For all
parameter values satisfying the above
condition $\xi_+$ is linearly unstable, and $\xi_-$ linearly unstable.

Let then ${\cal B}_\delta(\xi_\pm)$ be a disk of radius $\delta$ centered at
$\xi_\pm$ in the strip
$(0,\infty)\times [-\pi,\pi]$ of the $(w,\Delta)$-plane, and let ${\cal
B}_\delta^c(\xi_\pm)$
be its complement. Since the integral over ${\cal B}_\delta^c(\xi_\pm)$ of the
second set of terms in (\ref{linP-eq})
vanishes identically, it follows that the perturbation term $\int_{{\cal
B}_\delta^c(\xi_\pm)} d\xi \,\,\delta P(\xi)$
satisfies the same equation  as it would under the flow of the vector field in
(\ref{Zpm-eq})-(\ref{Wpm-eq}).
In that case, for the $+$ sign,  $\int_{{\cal B}_\delta^c(\xi_+)} d\xi
\,\,\delta P(\xi)$ increases in time, whereas
for the $-$ sign $\int_{{\cal B}_\delta^c(\xi_-)} d\xi \,\,\delta P(\xi)$
decreases in time, for any $\delta>0$. We therefore
see that solution $P_+(\xi)$ is linearly unstable, whereas the solution
$P_-(\xi)$ is linearly stable. In fact,
a direct integration of the full nonlinear equation (\ref{P-eq}) shows that
$P_-(\xi)$ is the global attractor for all
initial probability density functions $P(\xi)$. See \cite{ECC}.

This argument, given here assuming perfect independence of ``spins'' at
different shellnumbers, can
be generalized assuming a correlation of finite range $r$. In that case, one
can develop a similar
equation for the $r$-spin distribution $P_r(\xi_n,\xi_{n+1},...,\xi_{n+r})$. It
is found in the same
manner that a delta-function at the K41 fixed point value $\xi_-$ in all shells
is the unique, global
attracting solution. For details, see \cite{ECC}.

\subsection{Markov Chain Models of Multipliers}

In this second appendix we briefly discuss Markov chain models for the
multipliers.
We shall assume that $\xi_n=(w_n,\Delta_n)$ has statistics derived from a
stationary Markov
chain with single-shell distribution $P(\xi_n)$ and forward transition
probability
$T(\xi_{n+1}|\xi_n),\,\,n=1,2,3,...$ Because of the stationarity assumption,
these
functions do not depend upon shellnumber $n$. We first discuss the relation
between
the structure-function scaling exponents and the eigenvalues of certain
``transfer matrices''.
This is a particular example of the connection discussed in the text between
scaling
exponents and free energy functions. We next discuss the scaling properties of
the
mean fluxes of conserved quantities. In particular, we show how Markov chain
models
can yield a joint cascade of both energy and helicity.

To calculate the absolute structure functions, it is easiest to consider the
time-reversed Markov chain, with backward transition probability
$$ \widetilde{T}(\xi_n|\xi_{n+1}) =
{{T(\xi_{n+1}|\xi_n)P(\xi_n)}\over{P(\xi_{n+1})}}. $$
Then, using $\rho_n^p=\prod_{k=1}^n w_k^p=\prod_{k=1}^n e^{p\sigma_k}$,
\begin{eqnarray}
\langle \rho_n^p\rangle & = & \int d\xi_{n+1}\,\,P(\xi_{n+1})
                          \int d\xi_{n}
\,\,e^{p\sigma_n}\widetilde{T}(\xi_n|\xi_{n+1})
                   \cdots \int d\xi_{1}
\,\,e^{p\sigma_1}\widetilde{T}(\xi_1|\xi_2) \cr
\, & = &  \int d\xi_{n+1} \,\,P(\xi_{n+1}) \int d\xi_{1} \,\,
T_{(p)}^n(\xi_1|\xi_{n+1})
\end{eqnarray}
where the ``transfer matrix'' $T_{(p)}$ is defined by
$$ T_{(p)}(\xi_n|\xi_{n+1}) := e^{p\sigma_n} \widetilde{T}(\xi_n|\xi_{n+1}) $$
and $T_{(p)}^n$ is its $n$-fold convolution or matrix product. For large $n$,
we have asymptotically that
\be T_{(p)}^n(\xi_1|\xi_{n+1}) \sim \mu_{(p)}^n
R_{(p)}(\xi_1)L_{(p)}(\xi_{n+1})
    \lb{princ} \ee
where $\mu_{(p)}$ is the principal eigenvalue of $T_{(p)}$ and
$R_{(p)},L_{(p)}$
are corresponding right and left eigenfunctions. By the Perron-Frobenius
theorem, these
eigenvalues are real and non-negative, and the left and right eigenfunctions
may also
be chosen to be non-negative. Thus, we find that
$$ \langle \rho_n^p\rangle \sim \mu_{(p)}^n \langle L_{(p)}\rangle
\overline{R}_{(p)} $$
for $n\rightarrow\infty$, where $\langle L_{(p)}\rangle=\int d\xi \,\,P(\xi)
L_{(p)}(\xi)$
and $\overline{R}_{(p)} = \int d\xi \,\,R_{(p)}(\xi)$. In this way, we obtain
the relationship between the scaling exponents of structure functions and
eigenvalues
of the transfer matrices as
$$ \zeta_p = {{p}\over{3}} - \log_\lambda (\mu_{(p)}). $$
Exact moment constraints require that $\mu_{(3)}=1$, so that $\zeta_3=1$ within
a
Markov chain model. For details, see \cite{ECC}.

We now consider the asymptotics of mean fluxes of the inviscid invariants. In
the SABRA
model, the energy flux is given by
$$ \Pi_n^E = -a k_n {\cal T}_{n+1} + c k_{n-1} {\cal T}_n $$
and the helicity flux by
$$ \Pi_n^H = -2a \left({{a}\over{c}}\right)^n (k_n {\cal T}_{n+1} -k_{n-1}
{\cal T}_n). $$
Here ${\cal T}_n$ is a triple velocity product
$$ {\cal T}_n = {\rm Im}(u_{n+1}^*u_nu_{n-1}) =
{{1}\over{k_n}}\rho_{n+1}\rho_n\rho_{n-1}\sin\Delta_{n+1} $$
In terms of the scale-local variables $w_n,\Delta_n,\,\,n=1,2,3,...$
$$ \Pi_n^E = {{1}\over{\lambda}} \left[-a\cdot w_{n+2}\sin\Delta_{n+2} \cdot
w_{n+1}^2\cdot w_n^3
                + c\cdot w_{n+1}\sin\Delta_{n+1}\cdot w_n^2\right]
\prod_{k=1}^{n-1} w_k^3 $$
and
$$ \Pi_n^H = {{-2a}\over{\lambda}}\left({{a}\over{c}}\right)^n
              \left[w_{n+2}\sin\Delta_{n+2} \cdot w_{n+1}^2\cdot w_n^3
                   -w_{n+1}\sin\Delta_{n+1}\cdot w_n^2\right] \prod_{k=1}^{n-1}
w_k^3. $$

We now evaluate the mean values of these flux variables. If
$P_2(\xi_{n+2},\xi_{n+1})=
\widetilde{T}(\xi_{n+1}|\xi_{n+2})P(\xi_{n+2})$ is the joint distribution of
``spins''
$\xi_{n+2},\xi_{n+1}$ at two successive shells, then
\begin{eqnarray}
\, & & \langle \Pi_n^E\rangle = {{1}\over{\lambda}} \cr
\, & & \left[ -a \int d\xi_{n+2} \int d\xi_{n+1} \,\, P_2(\xi_{n+2},\xi_{n+1})
      \cdot w_{n+2}\sin\Delta_{n+2} \cdot w_{n+1}^2 \cdot \int d\xi_1\,\,
T_{(3)}^n(\xi_1|\xi_{n+1}) \right. \cr
\, & &  \left. +c \int d\xi_{n+1} \int d\xi_n \,\, P_2(\xi_{n+1},\xi_n)
      \cdot w_{n+1}\sin\Delta_{n+1}\cdot w_n^2 \int d\xi_1
\,\,T_{(3)}^{n-1}(\xi_1|\xi_{n}) \right]
\end{eqnarray}
and
\begin{eqnarray}
\, & & \langle \Pi_n^H\rangle = -{{2a}\over{\lambda}}
\left({{a}\over{c}}\right)^n \cr
\, & & \left[ \int d\xi_{n+2} \int d\xi_{n+1} \,\, P_2(\xi_{n+2},\xi_{n+1})
     \cdot w_{n+2}\sin\Delta_{n+2} \cdot w_{n+1}^2 \cdot \int d\xi_1\,\,
T_{(3)}^n(\xi_1|\xi_{n+1}) \right. \cr
\, & &  \left. -\int d\xi_{n+1} \int d\xi_n \,\, P_2(\xi_{n+1},\xi_n)
    \cdot w_{n+1}\sin\Delta_{n+1}\cdot w_n^2 \int d\xi_1
\,\,T_{(3)}^{n-1}(\xi_1|\xi_{n}) \right]
\end{eqnarray}

To calculate the mean energy flux asymptotically for large $n$, it suffices to
use the
previous asymptotic expansion (\ref{princ}) for $p=3$. This gives
$$ \langle \Pi_n^E\rangle \sim
-{{1}\over{\lambda}}(a\mu_{(3)}-c)\mu_{(3)}^{n-1}
           \langle w_2\sin\Delta_2\cdot w_1^2 L_{(3)}(w_1,\Delta_1) \rangle
\overline{R}_{(3)} $$
as $n\rightarrow \infty$. The only way that the energy flux can be
asymptotically constant, i.e.
independent of shellnumber $n$, is if $\mu_{(3)}=1$. This is another argument
for that constraint,
independent of that given in \cite{ECC}. In that case, we obtain finally that
$$ \langle \Pi_n^E\rangle \sim -{{1}\over{\lambda}}(a-c)
           \langle w_2\sin\Delta_2\cdot w_1^2 L_{(3)}(w_1,\Delta_1) \rangle
\overline{R}_{(3)} $$
asymptotically for $n\rightarrow \infty$.

However, for helicity flux, the contribution from the leading-order term in the
asymptotic
expansion (\ref{princ}) gives zero identically, because
$$ \langle w_{n+2}\sin\Delta_{n+2}\cdot w_{n+1}^2
L_{(3)}(w_{n+1},\Delta_{n+1}) \rangle=\langle w_{n+1}\sin\Delta_{n+1}\cdot
w_n^2 L_{(3)}(w_n,\Delta_n) \rangle, $$
by stationarity of the Markov chain. Thus, a non-vanishing contribution is
obtained only from the next order
in the asymptotic expansion for large $n$,
$$ T_{(3)}^n(\xi_n|\xi_{n+1}) \sim \mu_{(3)}^n R_{(3)}(\xi_1)L_{(3)}(\xi_{n+1})
             + \mu_{(3)}^{\prime n}
R_{(3)}^{\prime}(\xi_1)L_{(3)}^{\prime}(\xi_{n+1}) $$
where $\mu_{(3)}^{\prime}$ is the subleading eigenvalue of $T_{(3)}$ (i.e. the
complex
eigenvalue with next largest magnitude $|\mu_{(3)}^{\prime}|$ after
$|\mu_{(3)}|$), and
$R_{(3)}^{\prime},L_{(3)}^{\prime}$ the corresponding right and left
eigenfunctions. In that case,
$$ \langle \Pi_n^H\rangle \sim -{{2a}\over{\lambda}}
\left({{a}\over{c}}\mu_{(3)}^{\prime} \right)^n
     \left(1-{{1}\over{\mu_{(3)}^{\prime}}}\right)
     \langle w_2\sin\Delta_2\cdot w_1^2 L_{(3)}^{\prime}(w_1,\Delta_1) \rangle
\overline{R}_{(3)}^{\prime} $$
This flux is constant precisely when $\mu_{(3)}^{\prime}= c/a$. In that case,
$$ \langle \Pi_n^H\rangle \sim {{2a}\over{\lambda c}} (a-c)
   \langle w_2\sin\Delta_2\cdot w_1^2 L_{(3)}^{\prime}(w_1,\Delta_1) \rangle
\overline{R}_{(3)}^{\prime} $$
asymptotically for $n\rightarrow \infty$.

When $c/a<0$, a non-zero helicity flux is ruled out in an independent
multiplier model by the realizability
inequality $\langle w^3 \rangle>0$. However, in a Markov chain model, there is
no such constraint, because
the subleading eigenvalue $\mu_{(3)}^{\prime}$ may easily be negative. As a
concrete example, consider
the Markov chain with single-shell distribution
$$ P(\xi_n) =  C e^{-\beta\cdot\sin\Delta_n} P_W(w_n), $$
where $P_W(w_n)$ is any density on the interval $(0,\infty)$, and with
transition probability density
$$ \widetilde{T}(\xi_n|\xi_{n+1}) = P(\xi_n) \cdot \left[1 +
\left({{c}\over{a}}\right)
                         {\rm sgn}(\cos\Delta_n){\rm
sgn}(\cos\Delta_{n+1})\right]. $$
The latter kernel is non-negative when $|c/a|<1$. In fact, it is a rank 2
operator and can be written as
$$ \widetilde{T}(\xi_n|\xi_{n+1}) = R(\xi_n) L(\xi_{n+1}) +
\left({{c}\over{a}}\right) R^\prime(\xi_n) L^\prime (\xi_{n+1}) $$
where
$$ R(\xi_n) = P(\xi_n),\,\,\,\, L(\xi_n) = 1, \,\,\,\,
   R^\prime(\xi_n) = P(\xi_n){\rm sgn}(\cos\Delta_n),\,\,\,\, L^\prime(\xi_n) =
{\rm sgn}(\cos\Delta_n). $$
It is easily checked that these four vectors are a bi-orthonormal set, and can
be completed to a
bi-orthonormal basis. It follows directly that
$$ \int d\xi_n \,\,\widetilde{T}(\xi_n|\xi_{n+1}) =1, $$
so that $\widetilde{T}$ is a transition probability density, as claimed. Also
$P$ is its invariant distribution, because
$$ \int d\xi_{n+1} \,\,\widetilde{T}(\xi_n|\xi_{n+1}) P(\xi_{n+1}) = P(\xi_n).
$$
Furthermore, if $\int dw \,\,w^3 P_W(w)=1$, then by construction the only two
non-vanishing eigenvalues
of $T_{(3)}(\xi_n|\xi_{n+1})= w_n^3 \widetilde{T}(\xi_n|\xi_{n+1})$ are
$\mu_{(3)}=1$ and $\mu_{(3)}^\prime= c/a$, with
$$ R_{(3)}(\xi_n) = w_n^3 P(\xi_n),\,\,\,\, L_{(3)}(\xi_n) = 1 $$
$$ R_{(3)}^\prime(\xi_n) = w_n^3 P(\xi_n){\rm sgn}(\cos\Delta_n),\,\,\,\,
L_{(3)}^\prime(\xi_n) = {\rm sgn}(\cos\Delta_n). $$
This example is not completely realistic as a statistical model for the shell
dynamics, because the amplitude multipliers
$w_n,\,\,n=1,2,3,...$ form an i.i.d. sequence. Hence, it is not consistent with
the long power-law tail $\sim w^{-3}$
which we have observed in the numerical simulation. It is also does not have a
non-vanishing flux of helicity
because, unfortunately, the expectations $\langle w_2\sin\Delta_2\cdot w_1^2
L_{(3)}^{\prime}(w_1,\Delta_1) \rangle
= \overline{R}_{(3)}^{\prime}=0$ in this model. However, it illustrates how it
is possible to get the subleading
eigenvalue $c/a<0$  within a realizable Markov chain model.

{\bf Acknowledgments.} We thank L. Biferale, U. Frisch, K. Khanin, C. Meneveau
and T. Spencer
for discussions. Simulations were performed on the cluster computer supported
by the NSF grant
CTS-0079674 at the Johns Hopkins University.


\end{document}